\documentstyle[12pt,twoside,fleqn,espcrc1,epsf]{article}
\newcommand{\beq}{\begin{equation}}
\newcommand{\eeq}{\end{equation}}
\newcommand{\beqa}{\begin{eqnarray}}
\newcommand{\eeqa}{\end{eqnarray}}

\newcommand{\AmS}{{\protect\the\textfont2
  A\kern-.1667em\lower.5ex\hbox{M}\kern-.125emS}}

\title{Hadron Structure in the Non--Perturbative Regime of QCD:\\
Isospin Symmetry and its Violation
\footnote{
Plenary talk presented at the International Conference {\it "Quark--Lepton Nuclear
Physics 1997"}, Osaka, Japan, May 1997. Work supported
    in part by Deutsche Forschungsgemeinschaft (grant ME 864-11/1).} 
 }
\author{Ulf-G. Mei{\ss}ner\address{Forschungszentrum J\"ulich, 
Institut f\"ur Kernphysik (Theorie), 
D-52425 J\"ulich, Germany}}
\begin{document}

\maketitle

\vspace{-5cm}

\hfill KFA-IKP(TH)-1997-08

\vspace{4.5cm}

\begin{abstract}
\noindent I discuss recent progress made in calculating electromagnetic
corrections in the framework of the effective field theory of QCD.
In the case of elastic pion--pion scattering, strong interaction predictions
have been worked out to two loop accuracy. I present first results for
the electromagnetic corrections in the case of neutral pions. Here, the
only sizeable effect comes from the charged to 
neutral pion mass difference. In the presence
of nucleons, isospin violation can be measured in threshold pion 
photoproduction. I review the present status of the theoretical predictions
and the experimental data. I argue that a deeper understanding of isospin
violation based on a more precise study of such reactions can be achieved.
\end{abstract}
\section{Introduction}  

\noindent QCD S--matrix elements and 
transition currents in the non--perturbative
(low energy) regime can be calculated  accurately by
means of effective field theory methods, i.e. in chiral perturbation theory.
The effective chiral Lagrangian, formulated in terms of the asymptotically
observed fields (pions, nucleons) chirally coupled to external sources,
admits an expansion in small external momenta and quark masses. The
main physics goals of these studies are 
\begin{itemize}

\item[$\star$] to pin down the value of the scalar quark condensate,
 
\item[$\star$] to determine the ratios of the light quark masses,

\item[$\star$] to test the chiral anomaly of QCD. 

\end{itemize}
\noindent In this talk, I concentrate on
the question of isospin symmetry violation. There are two major sources
leading to a departure from this symmetry. First, there are the strong
interaction effects related to the light quark mass differences.
It is believed that the
ratio $m_d / m_u \simeq 2$ (at a canonical renormalization scale of 
1~GeV)\cite{glpr}.
This seems to indicate that isospin should be badly violated. However,
the difference $m_d - m_u \ll 1\,$GeV, which is the typical
strong interaction scale, and is thus effectively masked in
strong interaction processes (with the exception of some reactions, which
to leading order are proportional to $m_d - m_u$ like e.g. $\eta\to 3\pi$). 
Second, there are electromagnetic (virtual photon) effects. 
To really understand the isospin violation
due to the strong interactions, one has to be able to calculate these
em effects precisely. In the following, I will discuss some progress
made in calculating these in the framework of chiral perturbation theory.

\section{Chiral perturbation theory}

\noindent In the sector of the three light quarks $u$, $d$ and $s$, 
QCD admits a global chiral symmetry softly broken by the quark mass term,
\begin{equation}
{\cal H}_{\rm QCD} = {\cal H}_{\rm QCD}^0  +  {\cal H}_{\rm QCD}^{\rm SB}
=  {\cal H}_{\rm QCD}^0  + 
 m_u \, \bar u u + m_d \, \bar d d + m_s \, \bar s s \,\, ,
\end{equation}
where "light" means that the current quark mass at a renormalization
scale of $\mu = 1\,$GeV can be treated as small compared to the
typical scale of chiral symmetry breaking, $\Lambda_\chi \simeq 4 \pi
F_\pi \simeq 1.2\,$GeV, with $F_\pi \simeq 93\,$MeV the pion decay
constant. The tool to investigate these issues is chiral perturbation
theory (CHPT). In CHPT, the basic degrees of freedom are
the Goldstone boson fields coupled to external sources and matter
fields, like e.g. the nucleons. QCD is mapped onto an effective
hadronic Lagrangian formulated in terms of these asymptotically observed 
fields.Any matrix element involving nucleons, pions, photons and so on can
be classified according to its {\it chiral dimension}, which counts the
number of external momenta, quark mass insertions and inverse powers
of heavy mass fields. Denoting these small parameters collectively as
$p$, CHPT allows for a systematic perturbative expansion in powers of
$p$, with the inclusion of loop graphs and local terms of higher
dimension. The latter are accompanied by a priori unknown coupling
constants, the so--called low--energy constants (LECs). This is the
so--called {\it chiral} expansion, which is nothing but an energy
expansion reminiscent of the ancient Euler--Heisenberg treatment of
light--by--light scattering in QED at photon energies much smaller
than the electron mass. In QCD, the equivalent heavy mass scale is essentially
set by the first non--Goldstone resonances, i.e. the $\rho , \omega$ mesons.
Symbolically, any matrix--element can be expanded as
\begin{equation}
{\cal M} = \sum_n \, (p/\Lambda_\chi)^n \, f_n (p/\Lambda_\chi, \, g_i,\, 
\lambda / \Lambda_\chi)
\,\, ,
\end{equation}
where $ g_i$ denotes the LECs, $\lambda$ the scale of dimensional 
regularization and the $f_n$ are functions of order one which also 
contain the so--called
chiral logarithms. The important observation is that chiral symmetry {\it
bounds} the values of the counting index $n$ from {\it below}. This dual
expansion in small momenta and quark masses can be mapped one--to--one onto
an expansion in powers of Goldstone boson loops, where an $N$--loop graph
is suppressed by powers of $p^{2N}$~\cite{wein79}. The leading terms are
in general tree graphs with lowest order insertions leading to the
celebrated current algebra results. Space does not allow for a more 
detailed discussion of the method. Some recent developments are
summarized e.g. in ref.\cite{ulfrev} (and references therein).

\section{Elastic pion--pion scattering} 

\noindent The purest reaction to test the spontaneous and explicit chiral
symmetry breaking of QCD is elastic pion--pion scattering. In the threshold
region, the scattering amplitude can be decomposed as
\begin{equation}
t_l^I = q^{2l}\,\biggl[ a_l^I + b_l^I \, q^2 + {\cal O}(q^4) \biggr]
\,\, ,
\end{equation}
where $l$ denotes the pion angular momentum, $I$ the total isospin of the
two--pion system and $q$ the cms momentum.
Of particular interest are the S--wave scattering lengths $a_0^{0,2}$
since they vanish in the chiral limit of zero quark masses. At present,
they have been worked out to two loops in the chiral expansion. Consider
e.g. $a_0^{0}$,
\begin{equation}
a_0^{0} 
= \frac{7M_\pi^2}{32 \pi F_\pi^2}\,\biggl\{ 1 + a_1 \, M_\pi^2 + 
 a_2 \, M_\pi^4 + {\cal O}( M_\pi^6)   \biggr\}
\,\, ,
\end{equation}
with $M_\pi$ the pion mass and the terms $\sim a_i$ contain, of course, 
chiral logs.  The first term in this series is the celebrated
current algebra result of Weinberg\cite{wein66}, the second and third one
are the one-- and two--loop corrections given in \cite{gl84} 
and  \cite{bcegs}, respectively. Numerically, the series converges,
\begin{equation}
a_0^{0} = 0.156 \cdot ( 1 + 0.28 + 0.11 + {\cal O}( M_\pi^6) )
\,\, .
\end{equation}
Since the pion mass difference is almost entirely of em origin and the
Weinberg term changes from 0.156 to 0.146 when one uses the neutral instead
of the charged pion mass, one expects the em corrections to be of the
same size than the strong two--loop contributions. It is thus mandatory
to systematically investigate them. A first step towards this was
performed in ref.\cite{mms}. There, we constructed the
next--to--leading order chiral pion Lagrangian involving virtual
photons (the SU(3) case was already worked out in \cite{ru}). 
For that, one has to assign a chiral dimension to the
electric charge. Based on the observation that $\alpha = e^2/ 4\pi \simeq
1/137 \simeq M_\pi^2 / (4 \pi F_\pi)^2$, it is natural to count $e$ as
a small momentum, ${\cal O}(e) \sim {\cal O}(p)$. 
The virtual photon Lagrangian then takes the form 
\begin{equation}
{\cal L}_{\rm eff}^\gamma = {\cal L}_{\rm kin}^\gamma + 
{\cal L}_{\rm gauge}^\gamma + {\cal L}_{2}^\gamma +{\cal L}_{4}^\gamma
+ \ldots 
\end{equation}
where the ellipsis denotes higher order terms and the form of 
${\cal L}_{\rm kin}^\gamma + {\cal L}_{\rm gauge}^\gamma$ is
standard. To lowest order, one can only construct a single term,
\begin{equation} \label{L2}
{\cal L}_{2}^\gamma = C \, \langle Q_R U Q_L U^\dagger \rangle \,\,\, ,
\end{equation}
where $Q_{L,R}$ are spurions which transform linearly under left and
right SU(2) transformations. At the end, one sets $Q_L = Q_R = Q =
e\, ( 3 \tau^3 + {\bf 1})/6.$ The LEC $C$ can be determined from the
pion mass difference, $(\delta M_\pi^2)_{\rm em} = 2e^2C/F_\pi^2$.
Notice that when extended to SU(3), this term naturally leads to
Dashen's theorem, $(\delta M_\pi^2)_{\rm em} =(\delta M_K^2)_{\rm em}$.
An interpretation of this term and the LEC $C$ in terms of resonance
exchange can be found in \cite{egpdr}.

At
next--to--leading order, one has in total 13 terms (including the ones
which are only needed for renormalization) accompanied by 
scale--dependent LECs called $k_i (\lambda )$. 
These terms and the corresponding $\beta$--functions for the
LECs are enumerated in \cite{mms}. As it is the case for the hadronic
LECs in the two--flavor case \cite{gl84}, one can introduce scale--independent 
couplings, called $\bar{k}_i$, via
\beq
k_i^r (\lambda ) = {\kappa_i \over 32 \pi^2}\,\,  \biggl[\,\bar{k}_i + \ln 
\frac{M_{\pi^0}^2}{\lambda^2}\,\biggr] \quad .
\eeq
Notice that one chooses the neutral pion mass as the reference
scale. This is natural since the neutral pion mass is almost entirely 
a hadronic effect, in contrast to the charged one. 

In \cite{mms}, the numerical consequences for the process $\pi^0 \pi^0
\to \pi^0 \pi^0$ were worked out. This involves the diagrams c), d)
and j) shown in Fig.~1. In the $\sigma$--model gauge,
the insertion of the dimension two operator
Eq.({\ref{L2}) leads only to terms quadratic in the pion fields and
thus can be entirely absorbed in a redefinition of the pion propogator 
\beq \label{prop}
\Delta_\pi^{ab} (\ell)
= {i \delta^{ab}\over [\ell^2 - M_{\pi^0}^2 
- \delta M^2  \, ( 1- \delta^{3a} \, )]} \,\, ,
\eeq
with $\ell$ the pion four--momentum  and $'a,b\,'$ isospin indices.
It is then straightforward to work out the em corrections to the S--wave
scattering length $a_0$ and effective range $b_0$. While $a_0$ only
changes by 5\% (which is still smaller than the hadronic two--loop
correction), $b_0$ increases by 36\%. This can be understood by looking
at the partial wave amplitude $t_0 = t_0^0/3 + 2t_0^2/3$ 
divided by the scattering length
shown in Fig.~1. Above the $\pi^0\pi^0$ threshold at $W_0 =
2M_{\pi^0}$ the charged pion threshold opens at $W_c = 2M_{\pi^+}$. This
leads to a pronounced {\it cusp} effect which is expected to scale as
$\sqrt{ M_{\pi^+}^2 -  M_{\pi^0}^2} /M_{\pi^+} \simeq 26$\%. Also
shown in \cite{mms} is that the effect of the terms $\sim {\bar k}_i$
(graph j) is completely absent in this channel. The dominant
isospin--violating effect appears to be given by the charged to
neutral pion mass difference.
\begin{figure}[h]
\vspace{8.cm}
\includegraphics{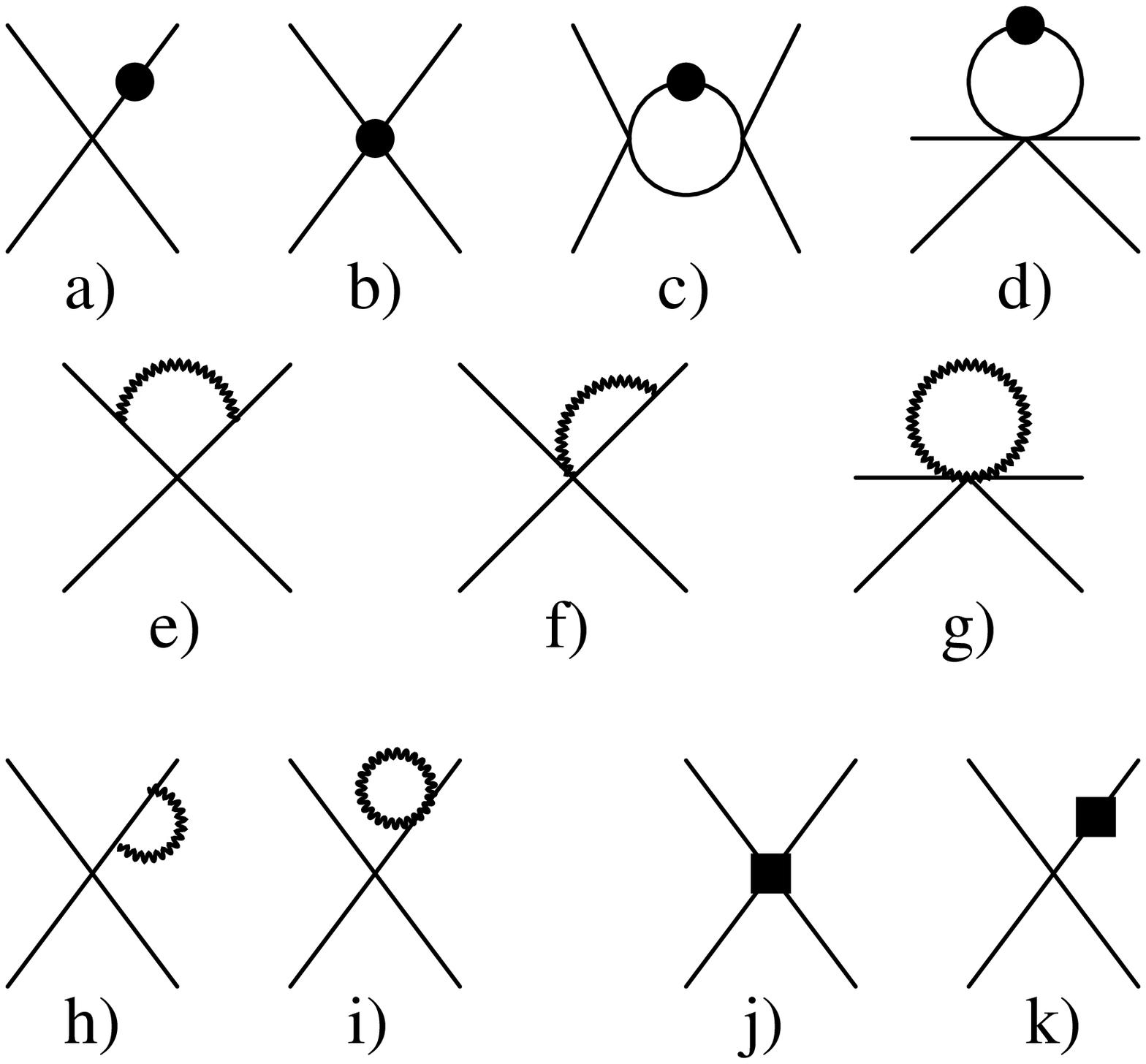}
\hspace{5.5cm}
\includegraphics{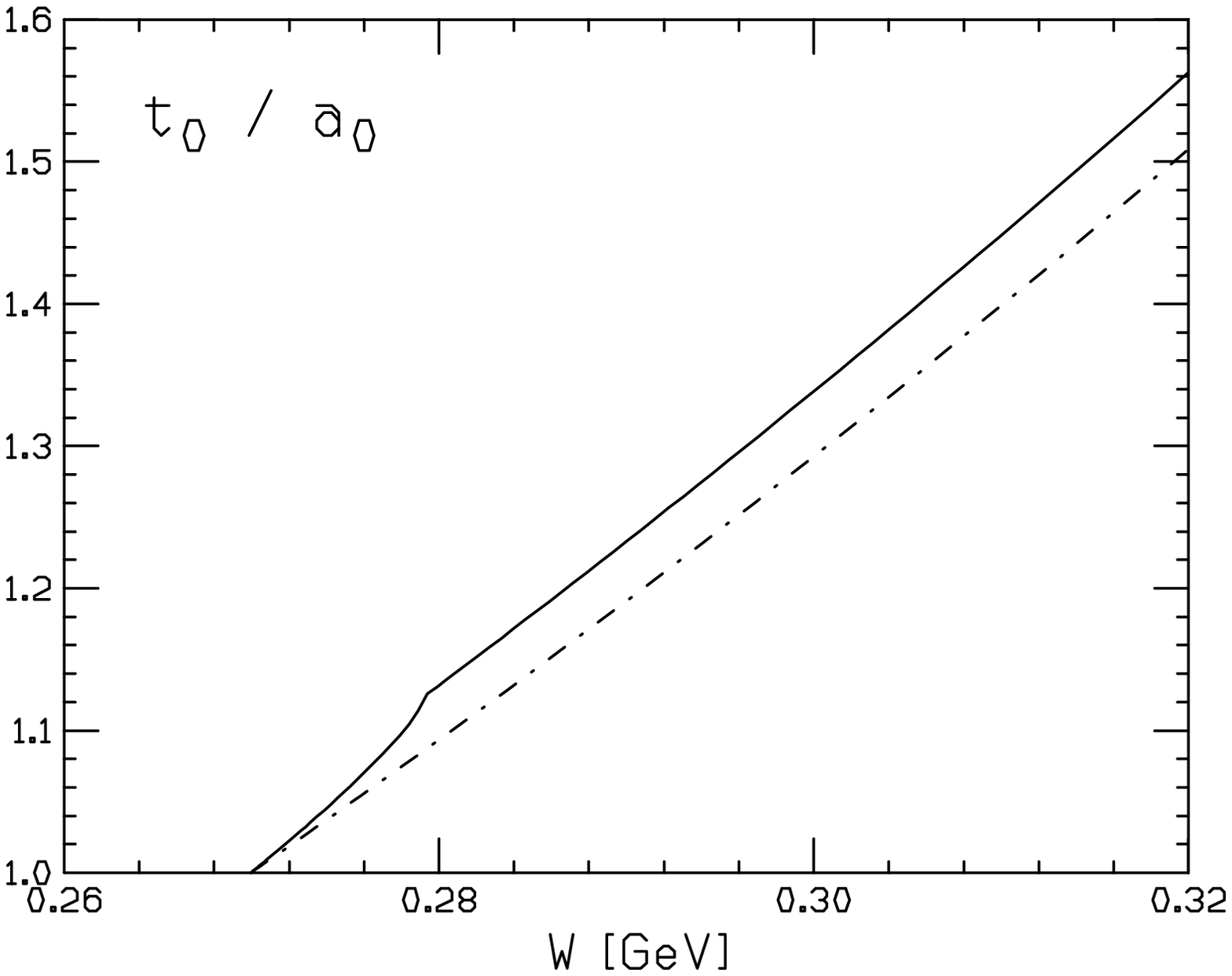}
\end{figure}

\vspace{-2.5cm}

\noindent Fig.~1:
Left panel: Tree and one--loop graphs contributing to the em
corrections for $\pi\pi$ scattering. Solid and wiggly lines denote
pions and photons, in order. The em contact insertions of dimension
two and four are depicted by filled circles and boxes, respectively.
Right panel: The normalized S--wave scattering amplitude for the
process $\pi^0 \pi^0 \to \pi^0 \pi^0$ versus the pion cm energy $W 
= \sqrt{s}$. The solid and dashed lines give
the results for $e \neq 0$ and $e = 0$, respectively.

\medskip

\noindent For reactions involving charged pions, matters are more
complicated. One finds e.g. that graph e) is infrared divergent, and this
IR divergence is not cancelled by the divergent parts of the
LECs. Indeed, one also has to account for the soft photon emission of
the external legs. Furthermore, to define the scattering lengths, one
has to properly subtract the long--range em interaction. A systematic
analysis of these effects is under way. It is important to note that
only in the effective field theory framework one can hope to develop a
consistent scheme to treat these various complications induced by the 
em interactions.

\section{The nucleon sector: General aspects}
 
\noindent As stressed  first by Weinberg \cite{weinmass}, reactions involving
nucleons and neutral pions are particularly suited to assess the 
quark mass difference $m_d - m_u$. In the pion sector, G--parity
allows to leading order only a term which is sensitive to the sum of
the quark masses, i.e. the terms $\sim m_d - m_u$ appear at
next--to--leading order. Furthermore, as discussed above, virtual
photons start to contribute already at lowest order. In the presence
of nucleons, the effective Lagrangian takes the following form
\beq
{\cal L}_{\pi N} = {\cal L}_{\pi N}^{(1)} +  {\cal L}_{\pi N}^{(2)}
+ {\cal L}_{\pi N}^{(3)} +  {\cal L}_{\pi N}^{(4)} + \ldots \,\,\, .
\eeq 
Since the charge matrix always appears quadratically in physical
processes, the leading term ${\cal L}_{\pi N}^{(1)}$ can not be directly
affected by the inclusion of virtual photons. The only effect at this
order comes from the covariant derivative, $D_\mu = \partial_\mu + i
\, A_\mu \,Q$ which generates vertices to be used in the calculation
of photon loops. At next order (dimension two), one has exactly two
new terms that influence the pion--nucleon couplings. Furthermore,
at this order exactly one term $\sim m_d - m_u$ appears
\cite{weinmit}. The corresponding LECs are finite. The dimension three and four
Lagrangians have not yet been worked out in full detail. I will
therefore take a more pedestrian approach here and try to estimate
what size of isospin violation {\it apart} from the one induced by the
pion mass difference we can expect. For that, I will review the
theoretical and experimental status of threshold pion photoproduction
and draw some lessons from that. For other discussions of isospin
violation in the pion--nucleon system, see e.g. van~Kolck \cite{bira}
and Bernstein \cite{aron}.

\section{Threshold pion photoproduction}

\noindent In the physical basis, we have four reactions, two involving
neutral and two involving charged pions, i.e. $\gamma p \to \pi^+ n$,
$\gamma n \to \pi^- p$, $\gamma p \to \pi^0 p$ and $\gamma n \to \pi^0
n$. Working to first order in the em coupling, the corresponding
amplitudes can be expressed in terms of {\it three} isospin
amplitudes, commonly denoted as $A^{(0,\pm)}$, via
\beqa
&& A(\gamma p \to \pi^+ n ) = \sqrt{2} \, ( A^{(0)} + A^{(-)} ) \,\, ,
\quad  A(\gamma n \to \pi^- p ) = \sqrt{2} \, ( A^{(0)} -A^{(-)} )
\,\, , \nonumber \\
&& A(\gamma p \to \pi^0 p ) \,\, =  A^{(0)} + A^{(+)} \,\, , \,\quad\qquad
A(\gamma n \to \pi^0 n ) \, =  -A^{(0)} + A^{(+)} \,\, .
\eeqa
If isospin were to be an {\it exact} symmetry, one would not need to
measure all four amplitudes but rather could deduce the fourth one 
from the ``triangle relation'', e.g.
\beq \label{tri}
A(\pi^0 n) = A (\pi^0 p) - {1\over \sqrt{2}} \, \biggl( A(\pi^+ n) 
+ A(\pi^- p) \biggr) \,\,\, .
\eeq
Any deviation from this is a measure of isospin violation either due
to the light quark mass difference or from the virtual photons. In the
threshold region, where the produced pion has a very small three
momentum, it is advantageous to decompose the amplitudes into S-- and 
P--wave multipoles. I will
now review the status of our knowledge about the (S--wave) electric
dipole amplitude $E_{0+}$ for the four reaction channels. Space
forbids to discuss in detail the interesting physics related to the
P--wave multipoles, in particular the novel low--energy theorems
for the multipole combinations $P_1$ and $P_2$
which have recently been derived \cite{bkmzpc}.

\subsection{Charged pion photoproduction}

\noindent Charged pion photoproduction at threshold is well described
in terms of the Kroll--Ruderman contact term, which is non--vanishing
in the chiral limit,
\beqa
E_{0+}^{\rm thr} (\pi^+ n) &=& \, \, \, \, 
\frac{e \, g_{\pi N}}{4 \pi \sqrt{2} m \, (1
  + \mu)^{3/2}} = \, \, \, \, \, 27.6 \cdot 10^{-3}/M_{\pi} \, \, 
\, , \nonumber \\
E_{0+}^{\rm thr} (\pi^- p) &=& -\frac{e \, g_{\pi N}}{4\pi \sqrt{2} m \, (1
  + \mu)^{1/2}}= -31.7 \cdot 10^{-3}/M_{\pi}  \, \, \, ,
\label{e0let}
\eeqa
with $\mu = M_{\pi}/m$ and using $g_{\pi N}^2/4 \pi = 14.28$, $e^2 /
4 \pi = 1 /137.036$,
$m=928.27\,$MeV and $M_{\pi}= 139.57\,$MeV. By now, 
all chiral corrections including the
third order in the pion mass have been calculated \cite{bkmcp}. The
chiral series is quickly converging and the theoretical error on the
CHPT predictions is rather small, see table~1. Notice that these
uncertainties do not account for the variations in pion--nucleon coupling constant,
about which no consensus has been reached yet. Also given in that
table are recent results from the dispersion theoretical (DR) analysis
of the Mainz group \cite{dht}. A theoretical uncertainty has not yet
been determined within that framework. The available threshold data
are quite  old, with
the exception of the recent TRIUMF experiment on the inverse reaction
$\pi^- p \to \gamma n$. While the overall agreement is quite good for 
the $\pi^+ n$ channel, in the $\pi^- p$ channel the CHPT prediction
is on the large side of the data. Clearly, we need more precise data
to draw a final conclusion. It is, however, remarkable to have
predictions with an error of only 2\%$\,$.
\vspace{-0.3cm}
\renewcommand{\arraystretch}{1.4}
\begin{table}[hbt]
\caption{Predictions and data for the charged pion electric dipole amplitudes.}
\label{tab:effluents}
\begin{tabular}{|l|c|c|c|}
\hline
                                & CHPT\protect{\cite{bkmcp}}
                                & DR\protect{\cite{dht}}     & Experiment  \\
\hline
$E_{0+}^{\rm thr} (\pi^+ n)$    & $28.2 \pm 0.6$ & $28.0$ 
                                & $27.9\pm 0.5$\protect{\cite{burg}},
                                $28.8 \pm 0.7$\protect{\cite{adam}} \\
$E_{0+}^{\rm thr} (\pi^- p)$    & $-32.7 \pm 0.6$ & $-31.7$ & $-31.4
                                \pm 1.3$\protect{\cite{burg}}, 
                               $-32.2 \pm 1.2$\protect{\cite{gold}}, 
                               $-31.5\pm 0.8$\protect{\cite{triumf}}  \\
\hline
\end{tabular}
\end{table}

\subsection{Neutral pion photoproduction off nucleons}

\noindent The threshold production of neutral pions is much more
subtle since the corresponding electric dipole amplitudes vanish in
the chiral limit. Space does not allow to tell the tale of the
experimental and theoretical developments concerning the electric
dipole amplitude for neutral pion production off protons, for details
see \cite{ulf95}. Even so the convergence for this particular
observable is slow, a CHPT calculation to order $p^4$ does allow to
understand the energy dependence of $E_{0+}$ in the threshold region
once three LECs are fitted to the total and differential cross section
data~\cite{bkme0p} as shown in fig.~\ref{fig:e0pp}. The threshold
value agrees with the data and the dispersion theoretical
determination, see table~\ref{tab:e0plus0}. More interesting is the
case of the neutron. Here, CHPT predicts a sizeably larger $E_{0+}$
than for the proton (in magnitude), whereas the dispersion relations
tend to give values of the same size (note however that the DR
treatment for the neutral channels is less stable than for the charged
ones). The CHPT prediction for $E_{0+} (\pi^0 n)$ in the threshold
region is shown in fig.~\ref{fig:e0pn}. Both amplitudes clearly
exhibit the unitary cusp due to the opening of the secondary
threshold, $\gamma p \to \pi^+ n \to \pi^0 p$ and $\gamma n \to \pi^-
p \to \pi^0 n$, respectively. Note, however, that while $E_{0+} (\pi^0
p)$ is almost vanishing after the secondary threshold, the neutron
electric dipole amplitude is sizeable ($-0.4$ compared to $2.8$ in
units of $10^{-3}/M_{\pi^+}$).

 
\renewcommand{\arraystretch}{1.4}
\begin{table}[hbt]
\caption{Predictions and data for the neutral pion electric dipole amplitudes.}
\label{tab:e0plus0}
\begin{tabular}{|l|c|c|c|}
\hline
                                & CHPT\protect{\cite{bkmcp}}
                                & DR\protect{\cite{dht}}     & Experiment  \\
\hline
$E_{0+}^{\rm thr} (\pi^0 p)$    & $-1.16$ & $-1.22$ 
                                & $-1.31\pm 0.08$\protect{\cite{fuchs}},
                                $-1.32\pm 0.11$\protect{\cite{berg}} \\
$E_{0+}^{\rm thr} (\pi^0 n)$    & $2.13$ & $1.19$ & 
                                $1.9 \pm 0.3$\protect{\cite{argan}} \\ 
\hline
\end{tabular}
\end{table}

\addtocounter{figure}{1}
\begin{figure}[t]
\begin{minipage}[ht]{77mm}
\vspace{0.5truecm}
\epsfxsize=7cm
\epsffile{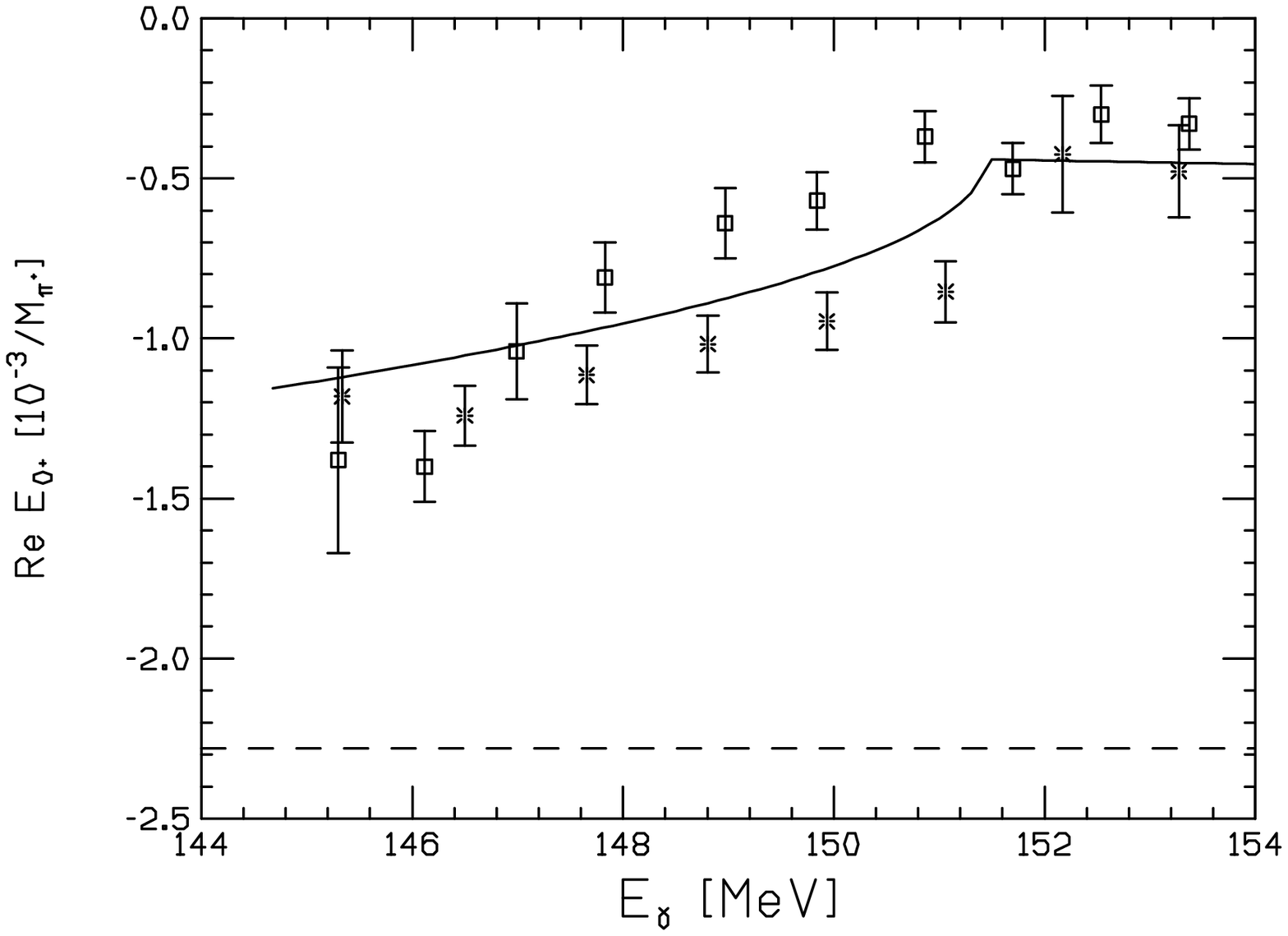}
\vspace{-0.5truecm}
\caption{CHPT prediction for the $\pi^0 p$ electric dipole amplitude
compared to the MAMI \protect{\cite{fuchs}} and SAL \protect{\cite{berg}} data.}
\label{fig:e0pp}
\end{minipage}
\hspace{1truecm}
\begin{minipage}[ht]{60mm}
\vspace{-.5truecm}
\epsfxsize=5.5cm
\epsffile{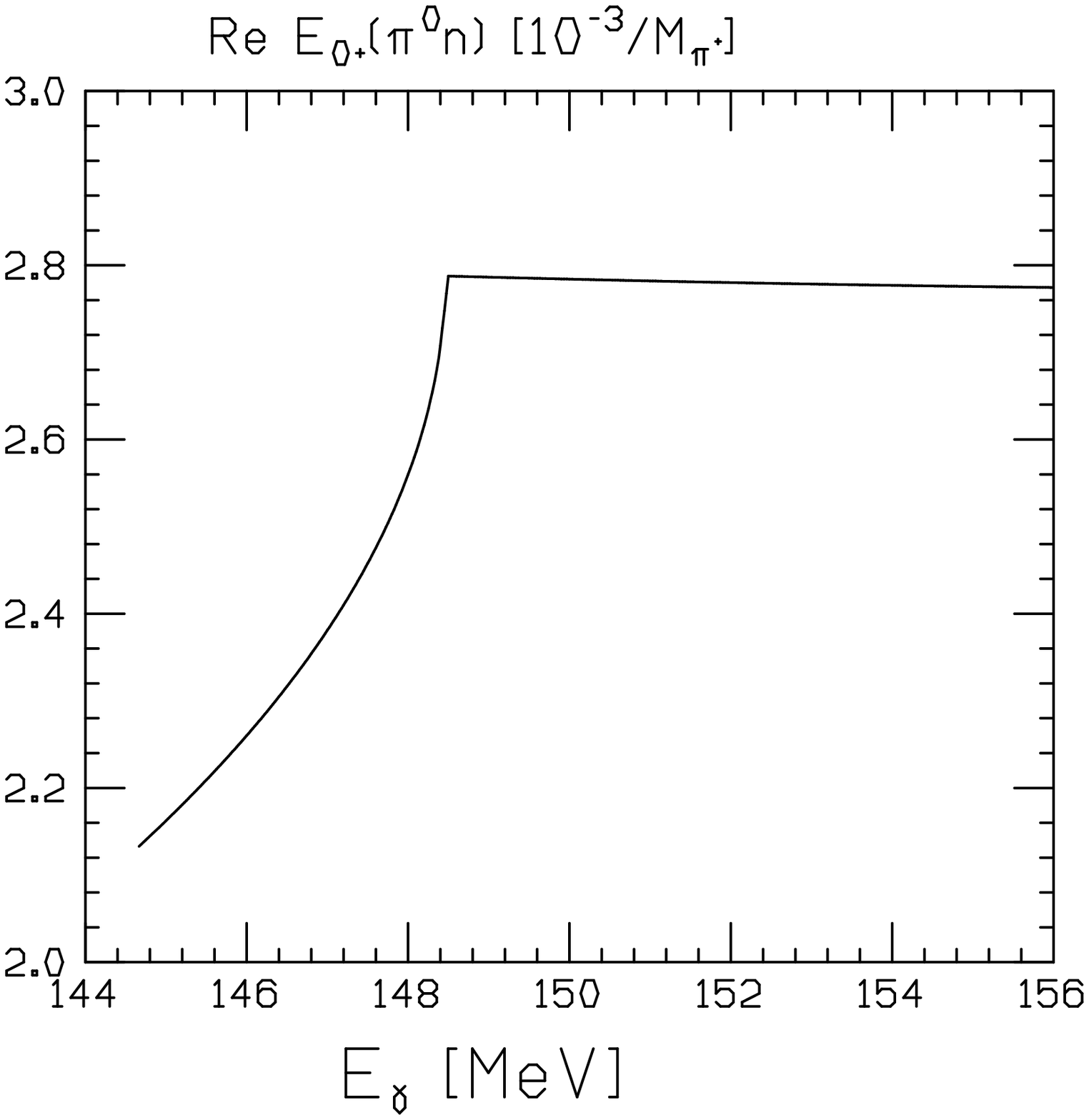}
\vspace{-1.5truecm}
\caption{CHPT prediction for the $\pi^0 n$ electric dipole amplitude.}
\label{fig:e0pn}
\end{minipage}
\end{figure}

\subsection{Neutral pion photoproduction off the deuteron}

\noindent The question arises how to measure the neutron amplitude?
The natural neutron target is the deuteron. The transition matrix for
$\pi^0$ production off the deuteron (d) takes the form
\beq
{\cal T} = 2i \, E_d \, {\vec J} \cdot \vec{\epsilon} + {\cal O}(\vec{q
  \,}) \,\,\, ,
\eeq
with ${\vec J}$ the total angular momentum of the d and
$\vec{\epsilon}$ the polarization vector of the photon. Although the
deuteron electric dipole amplitude could be calculated entirely within
CHPT, a more precise calculation is based on the approach suggested by
Weinberg~\cite{weinpid}, i.e. to calculate matrix elements of the type
$\langle \Psi_d | {\cal K} | \Psi_d\rangle$ by using deuteron wave
functions  $\Psi_d$ obtained from accurate  phenomenological NN potentials and
to chirally expand the kernel ${\cal K}$. Diagrammatically, one has
the single scattering (ss) terms which contain the desired $\pi^0 n$
amplitude. In addition, there are the so--called three--body (th)
contributions (meson exchange currents). To leading order $p^3$, one only
has the photon coupling to the pion in flight and the seagull
term~\cite{blvk}. The latter involves the charge exchange amplitude and is thus
expected to dominate the single scattering contribution. However, to
obtain the  same accuracy as for the ss terms, one has to calculate
also the corrections at order $p^4$. This has been done in
\cite{bblmvk}. It was shown that the next--to--leading order
three--body corrections and the possible four--fermion contact terms
do not induce any new unknown LEC and one therefore can calculate
$E_d$ in parameter--free manner. One finds
\beqa \label{Ed}
E_d &=& E_d^{\rm ss} + E_d^{\rm tb,3} + E_d^{\rm tb,4} \nonumber \\
 &=& 0.36 -1.90
-0.25 = (-1.8 \pm 0.2) \cdot 10^{-3}/M_{\pi^+} \,\,\, .
\eeqa 
Some remarks concerning this result are in order. First, one finds
indeed that the tb contribution is bigger than the single scattering
one. However, the former can be calculated precisely, i.e. the first
corrections amount to a meager 13\%. This signals good convergence.
I remark that a recent claim about large higher order (unitarity)
corrections \cite{pw} needs to be quantified in a consistent CHPT calculation.
Second, the resulting $E_d$ is very sensitive to $E_{0+} (\pi^0 n)$.
If one were to set $E_{0+} (\pi^0 n)= 0$, $E_d$ changes to $-2.6 
\cdot 10^{-3}/M_{\pi^+}$, i.e. the threshold cross sections would
change by a factor of two. Note that the theoretical error given in
Eq.(\ref{Ed}) is an educated guess, see \cite{bblmvk}.
Third, the CHPT prediction nicely agrees
with the empirical value of $E_d^{\rm exp} = (-1.7 \pm 0.2)\cdot
10^{-3}/M_{\pi^+}$~\cite{argan}. This agreement might, however, be fortitious since
the extraction of the empirical number relies on the input from the
elementary proton amplitude   to fix a normalization constant.  The
TAPS collaboration intends to redo this measurement at MAMI.

\subsection{The size of isospin violation}

\noindent I can now give a very rough estimate for the effects of
isospin violation beyond the one from the charged to neutral pion mass
difference. For that, let us compare the neutron amplitude as
predicted by CHPT or DR (labelled "pre") and compare with the result 
of the "triangle relation", Eq.(\ref{tri}), labelled "tri". This gives
\beqa
{\rm CHPT} &:& E_{0+}^{\pi^0 n, {\rm pre}} = 2.13 \,\, , \quad
 E_{0+}^{\pi^0 n, {\rm tri}} = 2.06 \,\, , \nonumber \\
{\rm DR}   &:& E_{0+}^{\pi^0 n, {\rm pre}} = 1.19 \,\, , \quad
 E_{0+}^{\pi^0 n, {\rm tri}} = 1.38 \,\, .
\eeqa
This indicates that threshold pion photoproduction is sensitive to
isospin violation induced by the light quark mass difference $m_d -
m_u$ and virtual photon effects (besides the ones leading to
$M_{\pi^0} \neq M_{\pi^\pm}$) of the order of a few up to
15\%. Clearly, such an estimate should only be considered {\it
  indicative} since it is not based on a fully self--consistent
calculation. Also, the rather large discrepancy one obtains in the DR
approach might to some extent reflect the uncertainty of the method
used in the analysis. This remains to be clarified. However, it is
rather obvious that combining precise {\it calculations} with very
{\it accurate} measurements, one can obtain significant information
on the origin of isospin violation in the pion--nucleon system.  
This is a rather novel situation in nuclear physics, namely that such
seemingly old--fashioned reactions can and will be used to test in 
a {\it quantitaive} manner our understanding of certain aspects of the
QCD in the low energy domain, i.e. where the strong coupling constant
is really large.

\vfill \eject

\section{Summary and outlook}

\noindent As outlined before, there are many problems still to be
tackled to gain a deeper insight into the violation of isospin in low
energy hadronic reactions. Let me mention here only a few of these:
\begin{itemize}

\item[$\star$] The effects of virtual photons on $\pi\pi$ scattering
involving charged pions are in the process of being worked out to one
loop order. These effects are complementary to the strong one and two
loop corrections calculated so far but are needed for a precise
comparison with the $K_{\ell 4}$ data expected from DA$\Phi$NE or the
pionium ones from CERN.
 
\item[$\star$] The construction of the effective chiral pion--nucleon
Lagrangian to one loop is underway,
\beq
{\cal L}_{\rm eff} [ U, N , {\cal A}_\mu ] = {\cal L}^{(2)} [\ldots ]
+ {\cal L}^{(3)} [\ldots ] + {\cal L}^{(4)} [\ldots ]
 \,\,\, ,
\eeq
with $U$, $N$ and ${\cal A}_\mu$ parametrizing the pions, nucleons and
virtual photons, in order. As explained before, the effective
Lagrangian  includes also the strong operators $\sim m_d - m_u$.

\item[$\star$] The next step will then be to apply this machinery to
pion photoproduction and pion--nucleon scattering. Over the last few 
years, there have been various claims of large isospin--breaking
effects in low--energy $\pi$N--scattering. Only with the consistent
machinery of the effective chiral Lagrangian involving virtual
photons, we can hope to put such claims on firm grounds (or to dismiss
them).

\end{itemize}

\noindent I finally remark that 20 years have passed since Weinberg's
seminal paper~\cite{weinmass} and that only now the theoretical
machinery as well as the experimental methods allow us to address these
questions in a truely {\it quantitative} manner. I am hopeful that
within a few years from now further considerable progress will have
been made.

\bigskip\bigskip\bigskip

\section*{Acknowledgements}

\noindent It is a great pleasure to thank the organizers, in
particular Profs. Y. Mizuno and H. Toki, for their invitation and
kind hospitality.


\end{document}